\documentclass[useAMS,usenatbib]{mn2e}
\usepackage{graphicx}
\usepackage{times}
\usepackage{multirow}
\usepackage{natbib}
\title[High-Res Observations of Na~I~D in 14 SNe~Ia]
  {Multi-Epoch High-Spectral-Resolution Observations of Neutral Sodium in 14 Type Ia Supernovae\thanks{Partially based on observations made with ESO Telescopes at the La Silla Paranal Observatory, Chile, under programme ID 289.D-5023, 290.D-5010, 290.D-5023, 091.D-0780.}}
\author[A. Sternberg et al.]
{A.~Sternberg,$^{1,2}$\thanks{E-mail: asternberg@mpa-garching.mpg.de}A.~Gal-Yam,$^{3}$ J.~D.~Simon,$^{4}$ F.~Patat,$^{5}$ W.~Hillebrandt,$^{1}$ M.~M.~Phillips,$^{6}$
\newauthor
R.~J.~Foley,$^{7}$ I.~Thompson,$^{6}$ N.~Morrell,$^{6}$ L.~Chomiuk,$^{8}$ A.~M.~Soderberg,$^{9}$ D.~Yong,$^{10}$
\newauthor
A.~L.~Kraus,$^{12}$ G.~J.~Herczeg,$^{13}$ E.~Y.~Hsiao,$^{6}$ S.~Raskutti,$^{14}$ J.~G.~Cohen,$^{15}$ P.~A.~Mazzali,$^{1,16}$
\newauthor
and K.~Nomoto,$^{17}$ \\
{1} Max Planck Institute for Astrophysics, Karl Schwarzschild St. 1, 85741 Garching bei M\"{un}chen, Germany \\
{2} Minerva Fellow \\
{3} Benoziyo Center for Astrophysics, Faculty of Physics, Weizmann Institute of Science, Rehovot 76100, Israel \\
{4} Observatories of the Carnegie Institution for Science, 813 Santa Barbara St., Pasadena, CA 91101, USA \\
{5} European Southern Observatory, Karl Schwarzschild St. 2, 85748 Garching bei M\"{un}chen, Germany \\
{6} Carnegie Observatories, Las Campanas Observatory, Casilla 601, La Serena, Chile \\
{7} Department of Astronomy, University of Illinois Champagne-Urbana, MC-221, 1002 W. Green Street, Urbana, IL 61801, USA \\
{8} Department of Physics and Astronomy, Michigan State University, East Lansing, MI 48824, USA \\
{9} Harvard-Smithsonian Center for Astrophysics, 60 Garden Street, Cambridge, MA 02138, USA \\
{10} Research School of Astronomy and Astrophysics, Australian National University, Canberra, ACT 2611 Australia \\
{11} Institute of Astronomy, University of Hawaii, 2680 Woodlawn Drive, Honolulu, HI 96822, USA \\
{12} Department of Astronomy, The University of Texas at Austin, 
2515 Speedway, Stop C1400, Austin, Texas 78712-1205, USA \\
{13} Kavli Institute for Astronomy and Astrophysics, Peking University; Yi He Yuan Lu 5, Hai Dian Qu, Beijing 100871, China \\
{14} Department of Astrophysics, Princeton University, Princeton, NJ 08540, USA \\
{15} Department of Astrophysics, California Institute of Technology, MC 249-17, Pasadena, CA 91125, USA \\
{16} Astrophysics Research Institute, Liverpool John Moores University, Liverpool L3 5RF, United Kingdom \\
{17} Kavli Institute for the Physics and Mathematics of the Universe (WPI), The University of Tokyo, Kashiwa, Chiba 277-8583, Japan}
\date{Released 2013xxxx XX}

\pagerange{\pageref{firstpage}--\pageref{lastpage}} \pubyear{2013}

\def\LaTeX{L\kern-.36em\raise.3ex\hbox{a}\kern-.15em
    T\kern-.1667em\lower.7ex\hbox{E}\kern-.125emX}

\begin{document}

\label{firstpage}

\maketitle

\begin{abstract}
One of the main questions concerning Type Ia supernovae is the nature of the binary companion of the exploding white dwarf. A major discriminant between different suggested models is the presence and physical properties of circumstellar material at the time of explosion. If present, this material will be ionized by the ultra-violet radiation of the explosion and later recombine. This ionization-recombination should manifest itself as time-variable absorption features that can be detected via multi-epoch high-spectral-resolution observations. Previous studies have shown that the strongest effect is seen in the neutral sodium D lines. We report on observations of neutral sodium absorption features observed in multi-epoch high-resolution spectra of 14 Type Ia supernova events. This is the first multi-epoch high-resolution study to include multiple SNe. No variability in line strength that can be associated with circumstellar material is detected. We find that $\sim18\%$ of the events in the extended sample exhibit time-variable sodium features associated with circumstellar material. We explore the implication of this study on our understanding of the progenitor systems of Type Ia supernovae via the current Type Ia supernova multi-epoch high-spectral-resolution sample.
\end{abstract}


%=============================================
\section{INTRODUCTION}
\label{sec:intro}
%=============================================

Type Ia supernovae (SNe~Ia) are widely accepted to be the thermonuclear explosion of accreting carbon-oxygen white-dwarfs (WDs) in close binary systems \citep{HoyleFowler1960ApJ132_565}. Despite numerous studies the nature of the companion of the exploding WD still remains uncertain. Presently, there are two models that are widely accepted as plausible descriptions of SN~Ia progenitor systems. In the first model, the single-degenerate (SD) model, the companion is a non-degenerate star that transfers mass onto the WD \citep{WhelanIben1973ApJ186_1007}. In a second, the double-degenerate (DD) model, the secondary is a WD \citep{IbenTutukov1984ApJS54_335, Webbink1984ApJ277_355}. A third model, less widely accepted at present, is the core-degenerate (CD) model in which a WD merges with the core of an evolved star during, or shortly after, the common-envelope phase \citep{Sparks&Stecher1974ApJ188_149, LivioRiess2003ApJ594_93, KashiSoker2011MNRAS417_1466, IlkovSoker2012MNRAS419_1695}. One of the major differences between current alternative models is the properties of the gaseous environment surrounding the WD at the time of explosion. Therefore, detection of absorption by circumstellar material (CSM) in the spectra of SNe~Ia, or the lack thereof, and the study of its properties might help to disentangle the plausible from the implausible progenitor systems. \par

\citet[][hereafter P07]{Patat_et_al2007Sci317_924} were the first to report the detection of circumstellar material in a SN~Ia, SN~2006X, based on time-variability observed in the neutral sodium (Na I) D absorption lines (restframe wavelengths $ \lambda \lambda 5890, 5896$). The lack of time variability in the Ca~II~H~\&~K lines led P07 to conclude that the observed change in the sodium absorption features was due to the ionization and recombination of circumstellar material, and not a geometrical line-of-sight effect. P07 based their conclusion arguing that SN~Ia UV radiation is severely line-blocked by heavy elements bluewards of $3500\rm\AA$ \citep{Pauldrach1996AA312_525, Mazzali2000AA363_705}, and is  therefore capable of ionizing only material that is relatively near the explosion. If the material density is sufficiently high it will recombine after maximum light giving rise to variability in observed absorption features. They suggested a single-degenerate progenitor system for SN~2006X. Following P07 similar studies provided a mixture of results both of detection of time-variable absorption features associated with CSM - SN~2007le \citep{Simon_et_al2009ApJ702_1157}, SN~1999cl \citep[][using low-spectral-resolution]{Blondin_et_al2009ApJ693_207} and PTF~11kx \citep{Dilday_et_al2012Sci337_942} - and of non-detection of such features - SN~2000cx \citep{Patat_et_al2007AA474_931}, SN~2007af \citep{Simon_et_al2007ApJ671_25}. \citet{Patat_et_al2013A&A549_62} detected marginally significant time-variability in the sodium features of SN~2011fe that were anyway compatible with interstellar material (ISM) and not with CSM. This mixture of results suggests that SNe~Ia might actually arise from two types of progenitor systems, as is suggested by other studies \citep{MannucciDella-VallePangia2006MNRAS370_773, Sullivan_et_al2010MNRAS406_782, Wang_et_al2013Sci340_170, Maguire_et_al2013accepted}. The published sample of SNe with multiple high-resolution spectra is too small to allow for a significant conclusion to be drawn. Moreover, it also prevents us from deriving a robust ratio between SNe~Ia that show evidence of CSM and those that do not. For that, a larger sample of SN~Ia multi-epoch high-resolution spectra is needed. \par

\citet[][hereafter S11]{Sternberg_et_al2011Sci333_856} adopted a different approach based on a statistical analysis of a single-epoch high-spectral-resolution sample consisting of 35 SN~Ia events. S11 showed that SNe~Ia in nearby spiral hosts exhibit a statistically significant overabundance of features that are blueshifted relative to the strongest absorption feature, and this was interpreted as evidence of outflows from $\sim20-25\%$ of their progenitor systems. \citet{Foley_et_al2012ApJ_752_101} showed that events classified as blueshifted \'{a} la S11 tend to have higher ejecta velocities and redder colors at maximum light compared to the rest of the SN~Ia sample. \citet{Maguire_et_al2013accepted} used a classification scheme slightly different then that used by S11 to analyze an extended sample consisting of the S11 sample and an additional 17 intermediate-resolution single-epoch spectra, showing a $\sim20\%$ excess of events with blueshifted features. These analyses lend strong support to the existence of CSM in a non-negligible fraction of SNe~Ia and the possibility of a bimodal distribution of SN~Ia progenitors. Nevertheless, one cannot use the S11 sample or the \citeauthor{Maguire_et_al2013accepted} sample to study the properties of the CSM as it is not possible to confidently distinguish between CSM and ISM features for an individual SN in single-epoch spectra. Multi-epoch observations are needed. \par

In this paper we present multi-epoch high-spectral-resolution observations of 14 Type Ia events, some of which are from the S11 sample for which only single-epoch spectra were previously published. Publication of such data will help make the published SN~Ia multi-epoch high-spectral-resolution sample more complete. We use this data set, plus literature observations of six other SNe, to provide the first robust estimate of the fraction of SNe Ia exhibiting variable absorption. We discuss the robustness of the time-variability non-detection and how this effects the consistency of the current multi-epoch high-spectral-resolution sample with previous studies. \par

% =============================================
\section{DATA}
\label{sec:data}
% =============================================

Observations were performed using either the Ultraviolet and Visual Echelle Spectrograph \citep[UVES;][]{Dekker_et_al2000SPIE4008_534D} mounted on the Very Large Telescope array (VLT) UT2, the HIgh Resolution Echelle Spectrograph \citep[HIRES;][]{Vogt_et_al1994SPIE2198_362} mounted on the Keck I telescope, the Magellan Inamori Kyocera Echelle \citep[MIKE;][]{Bernstein_et_al2003ProcSPIE4841_1694} spectrograph mounted on the Magellan II Clay telescope, and the East Arm Echelle \citep[EAE;][]{LibbrechtPeri1995PASP107_62} spectrograph mounted on the 200 inch Hale telescope at Palomar. Information regarding the discovery and spectroscopic observations of individual SNe  can be found in appendix \ref{app:obs_desc}. UVES data were reduced using the latest ESO Reflex reduction pipeline. HIRES spectra were reduced using the MAuna Kea Echelle Extraction (MAKEE) data reduction package (written by T. Barlow; http://spider.ipac.caltech.edu/staff/tab/makee/). MIKE spectra were reduced using the latest version of the MIKE pipeline (written by D. Kelson; http://code.obs.carnegiescience.edu/mike/). EAE data were reduced using standard IRAF routines.

% =============================================
\section{RESULTS}
\label{sec:results}
% =============================================

\begin{figure}

  \includegraphics[width=1.0\columnwidth]{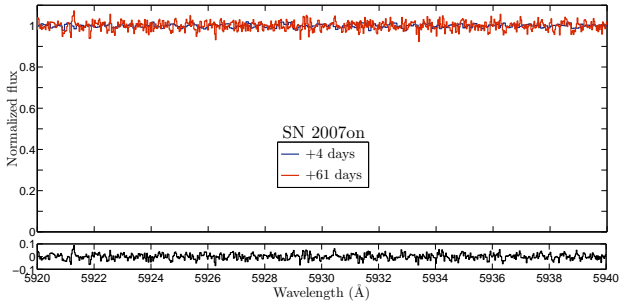}
  \caption{The dual-epoch spectra of SN~2007on. The first epoch is given in blue while the second is given in red. The difference spectrum (black line; in units of normalized flux) is given in the narrow panel at the bottom of the plot. Based on the host galaxy redshift, $z=0.006494$, the host/SN Na~I~D features are expected to appear around 5928\AA\, \& 5934\AA (for the D$_2$ and D$_1$, respectively). No features are observed.}
\label{fig:07on}
\end{figure}

\begin{figure}
  \includegraphics[width=1.0\columnwidth]{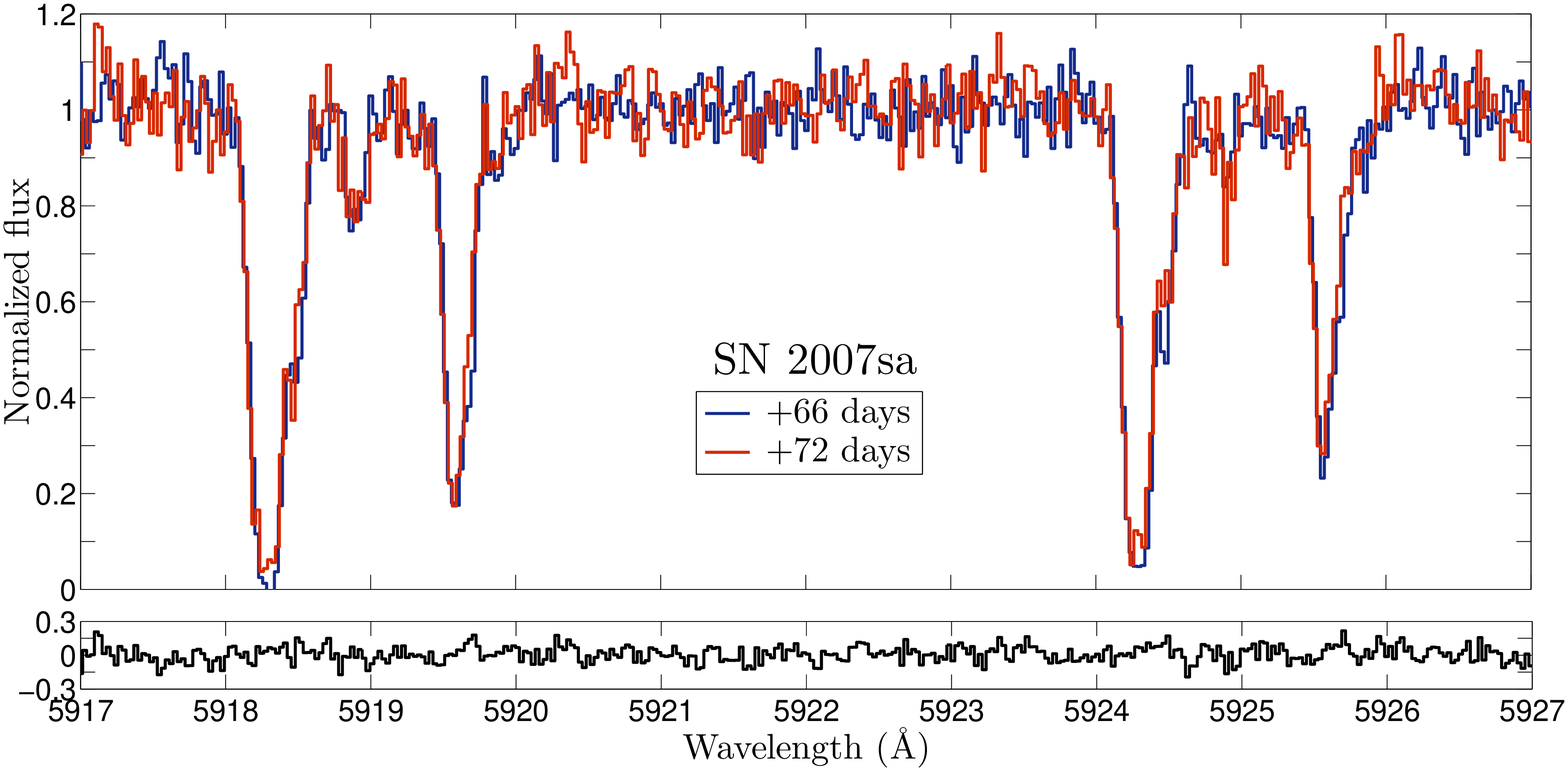}
  \caption{The dual-epoch spectra of SN~2007sa. The first epoch is given in blue while the second is given in red. The difference spectrum (black line) is given in the narrow panel at the bottom of the plot.}
\label{fig:07sa}
\end{figure}

\begin{figure}
  \includegraphics[width=1.0\columnwidth]{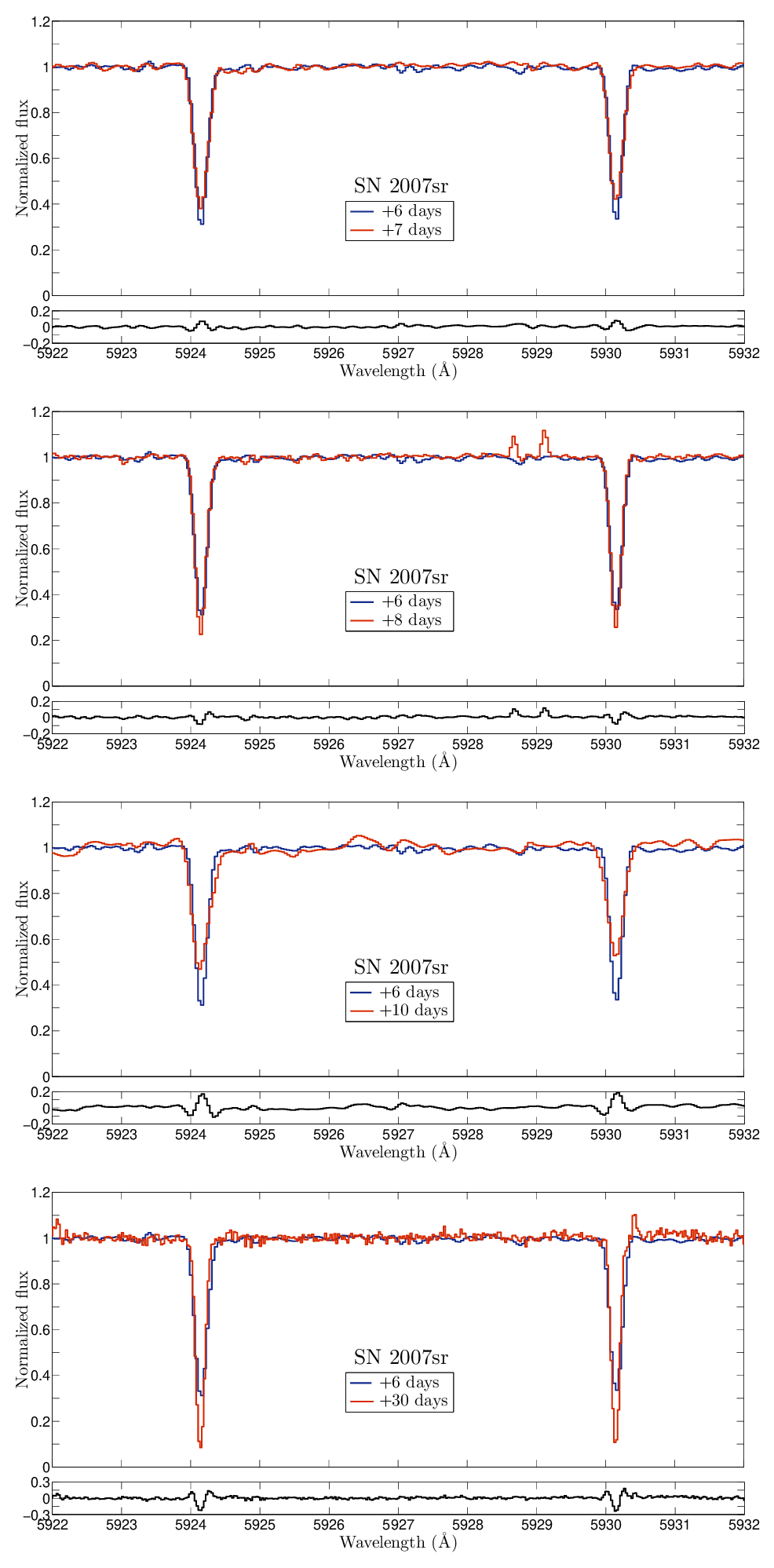}
  \caption{The multi-epoch spectra of SN~2007sr. Each plot shows a comparison of the first epoch (blue) and one of the other four epochs (red). The black lines given in the lower panel of each plot is the difference spectrum. As the measured EW of the different epochs are consistent with one another the observed difference in the line profiles are due to the different resolution of the spectrographs used on the different epochs.}
\label{fig:07sr}
\end{figure}

\begin{figure}
  \includegraphics[width=1.0\columnwidth]{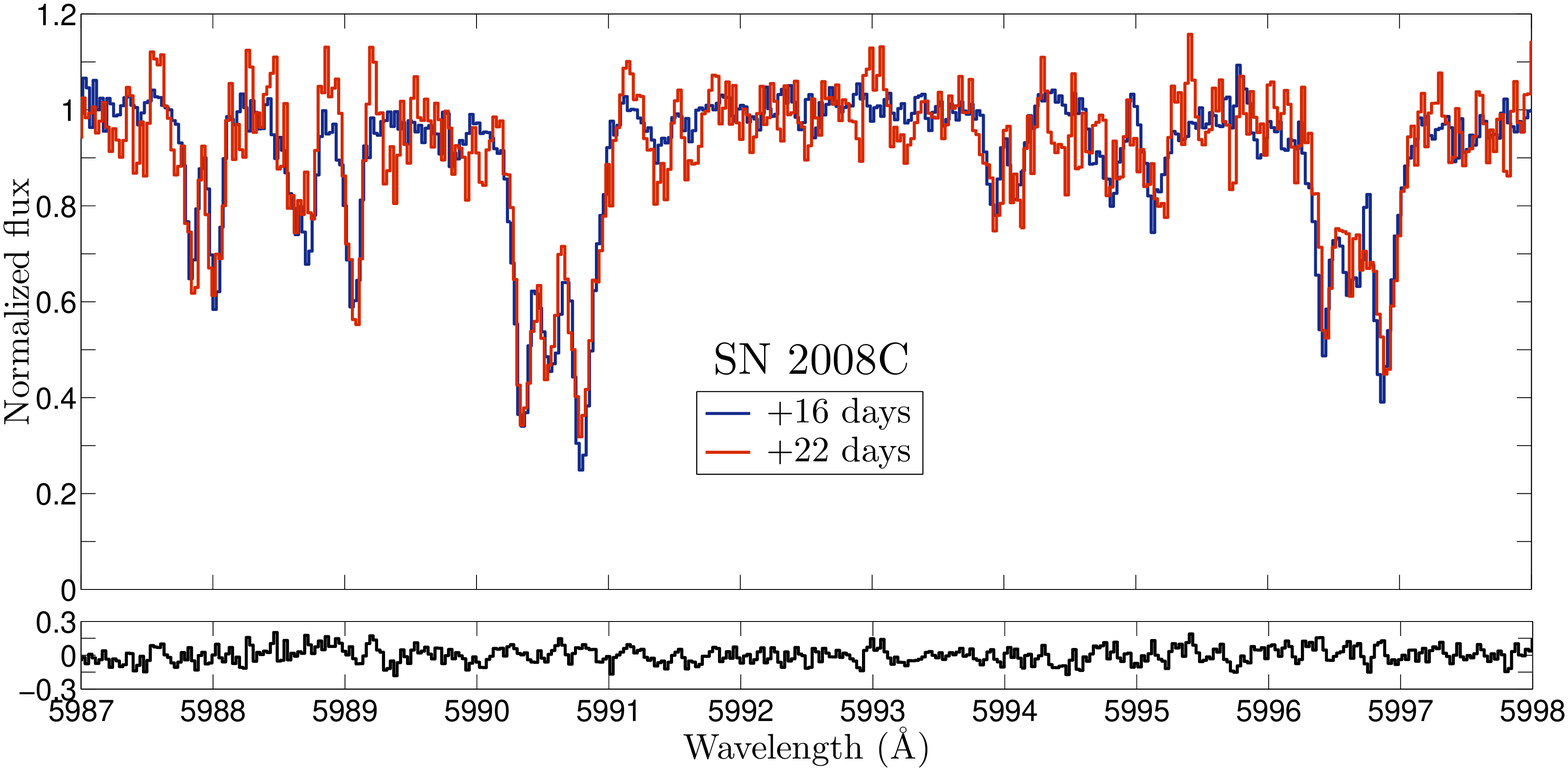}
  \caption{The dual-epoch spectra of SN~2008C. The first epoch is given in blue while the second is given in red. The difference spectrum (black line) is given in the narrow panel at the bottom of the plot.}
\label{fig:08C}
\end{figure}

\begin{figure}
  \includegraphics[width=1.0\columnwidth]{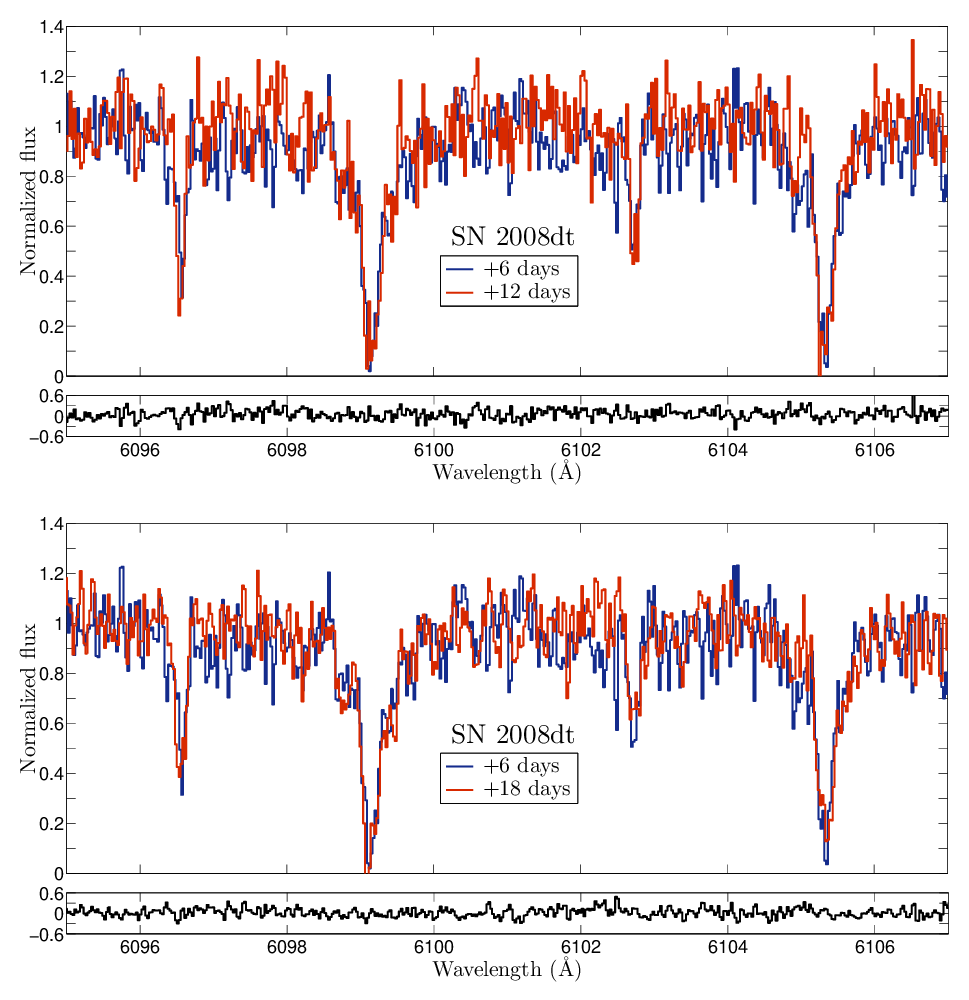}
  \caption{The triple-epoch spectra of SN~2008dt. Color code is the same as in previous figures. The upper panel shows the first and second epochs and the middle panel shows the first and third.}
\label{fig:08dt}
\end{figure}

\begin{figure}
  \includegraphics[width=1.0\columnwidth]{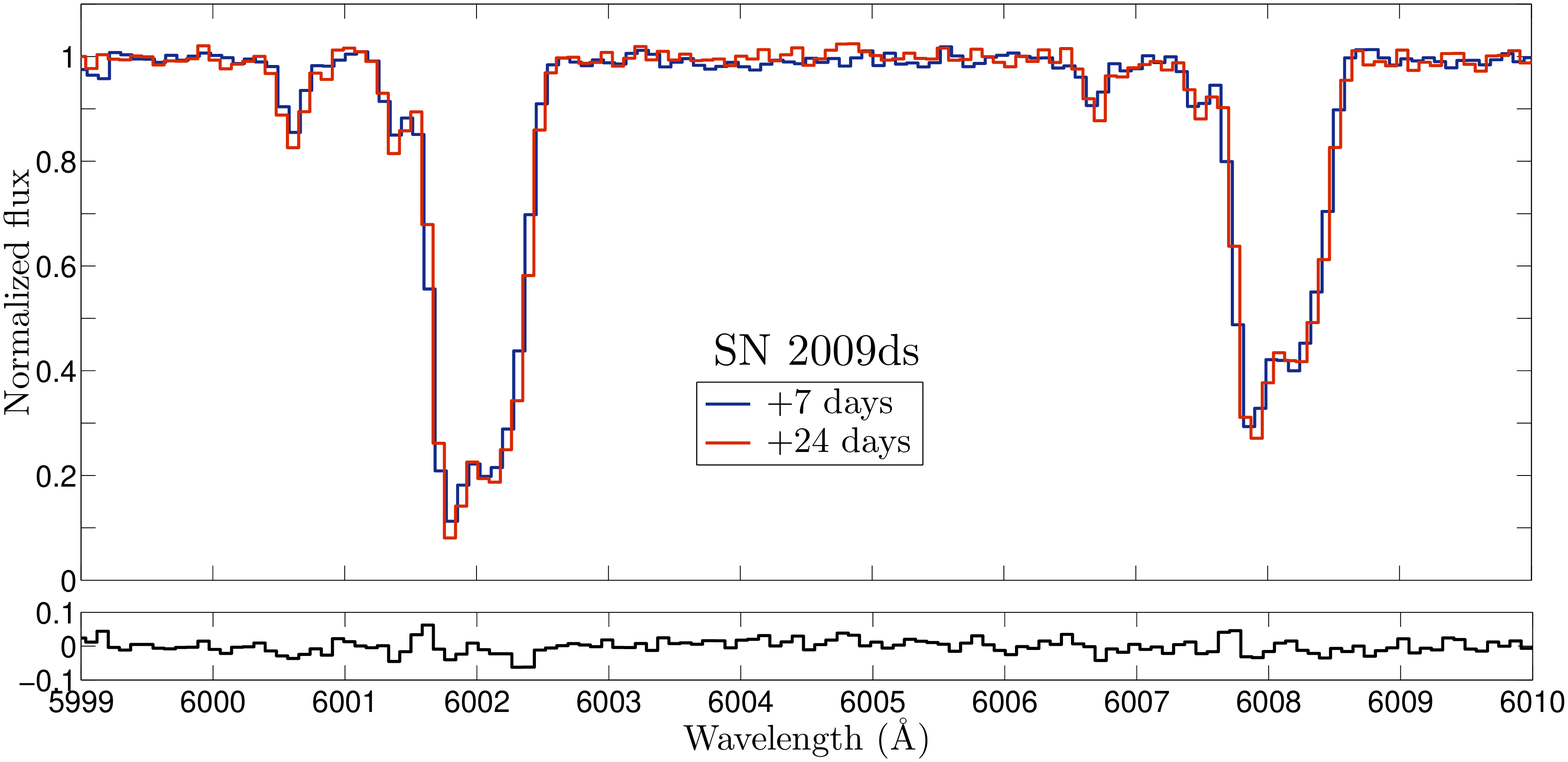}
  \caption{The dual-epoch spectra of SN~2009ds. Color code is the same as in the previous figures.}
\label{fig:09ds}
\end{figure}

\begin{figure}
  \includegraphics[width=1.0\columnwidth]{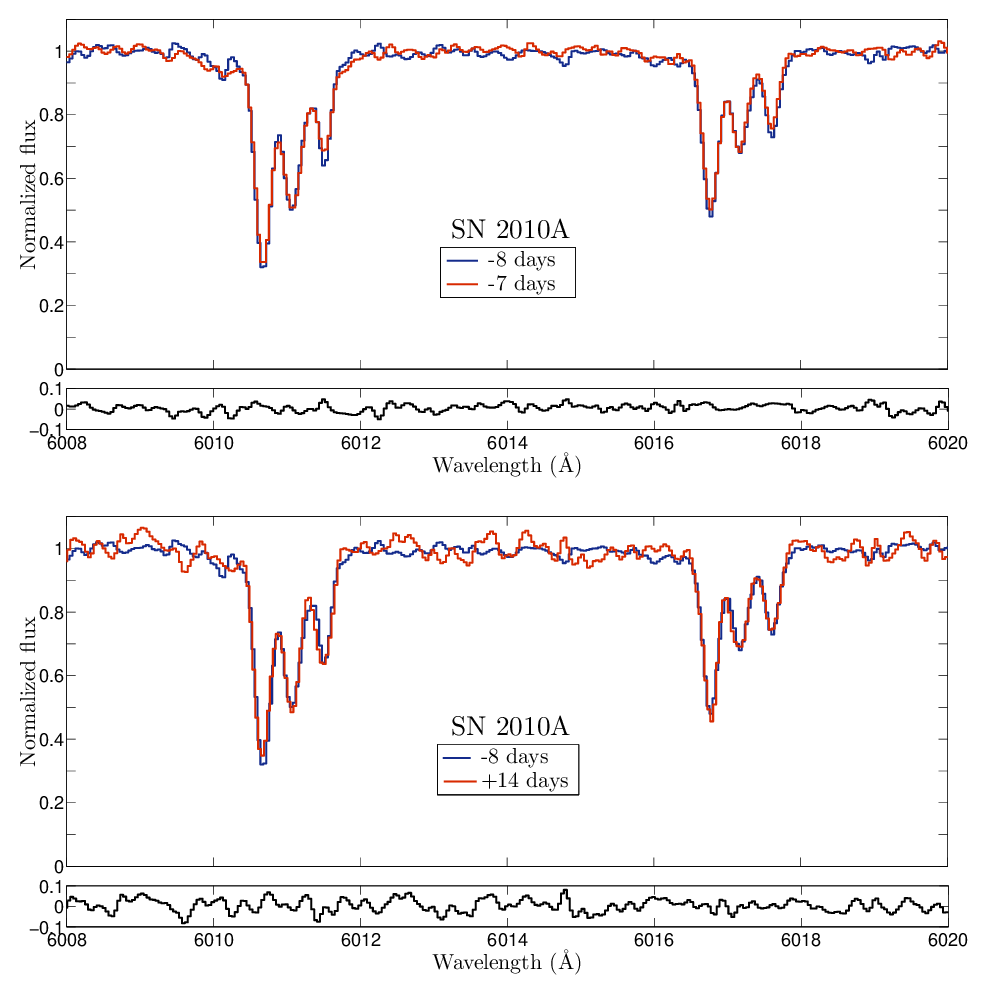}
  \caption{The triple-epoch spectra of SN~2010A.}
\label{fig:10A}
\end{figure}

\begin{figure}
  \includegraphics[width=1.0\columnwidth]{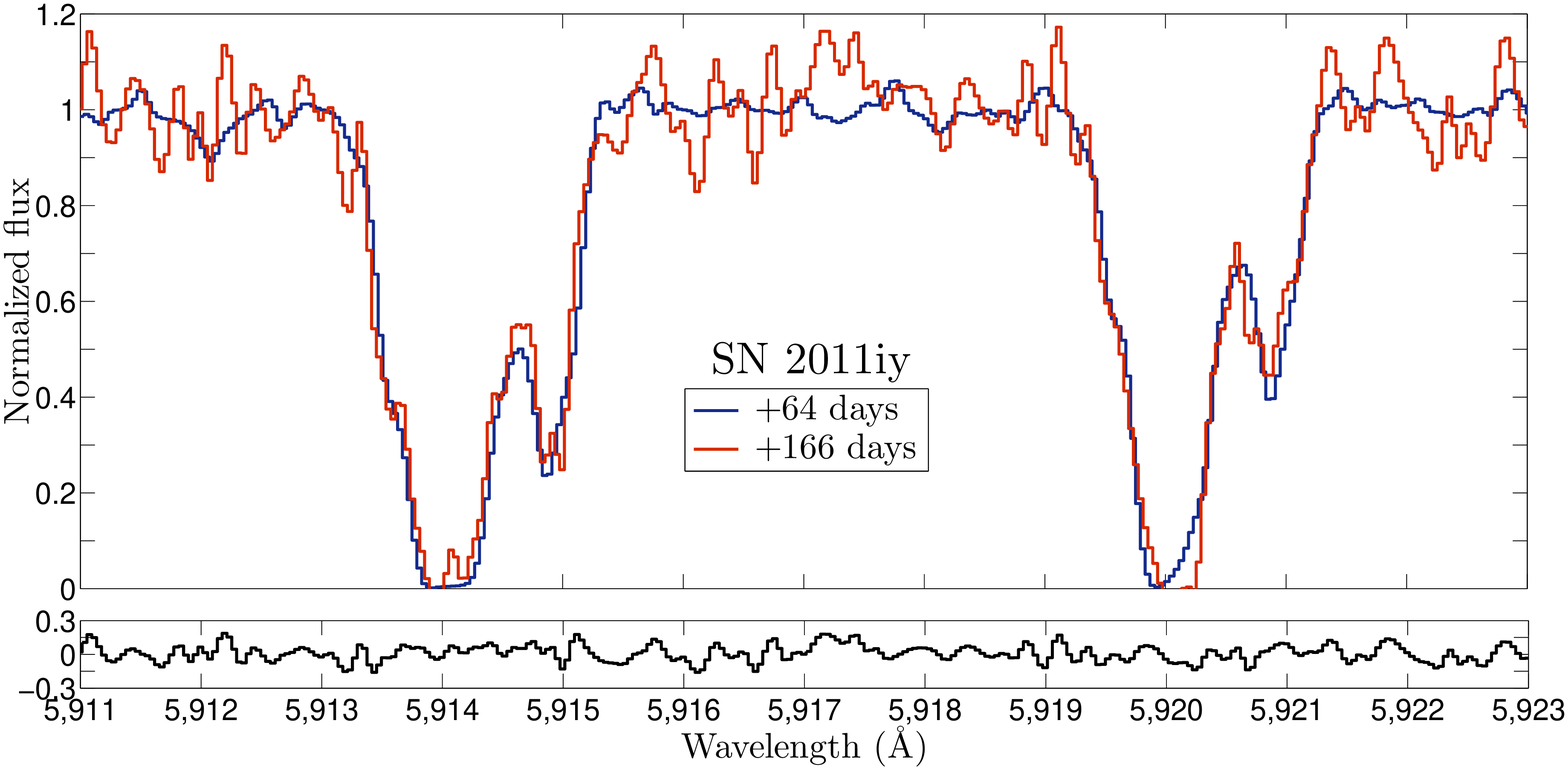}
  \caption{The dual-epoch spectra of SN~2011iy.}
\label{fig:11iy}
\end{figure}

\begin{figure}
  \includegraphics[width=1.0\columnwidth]{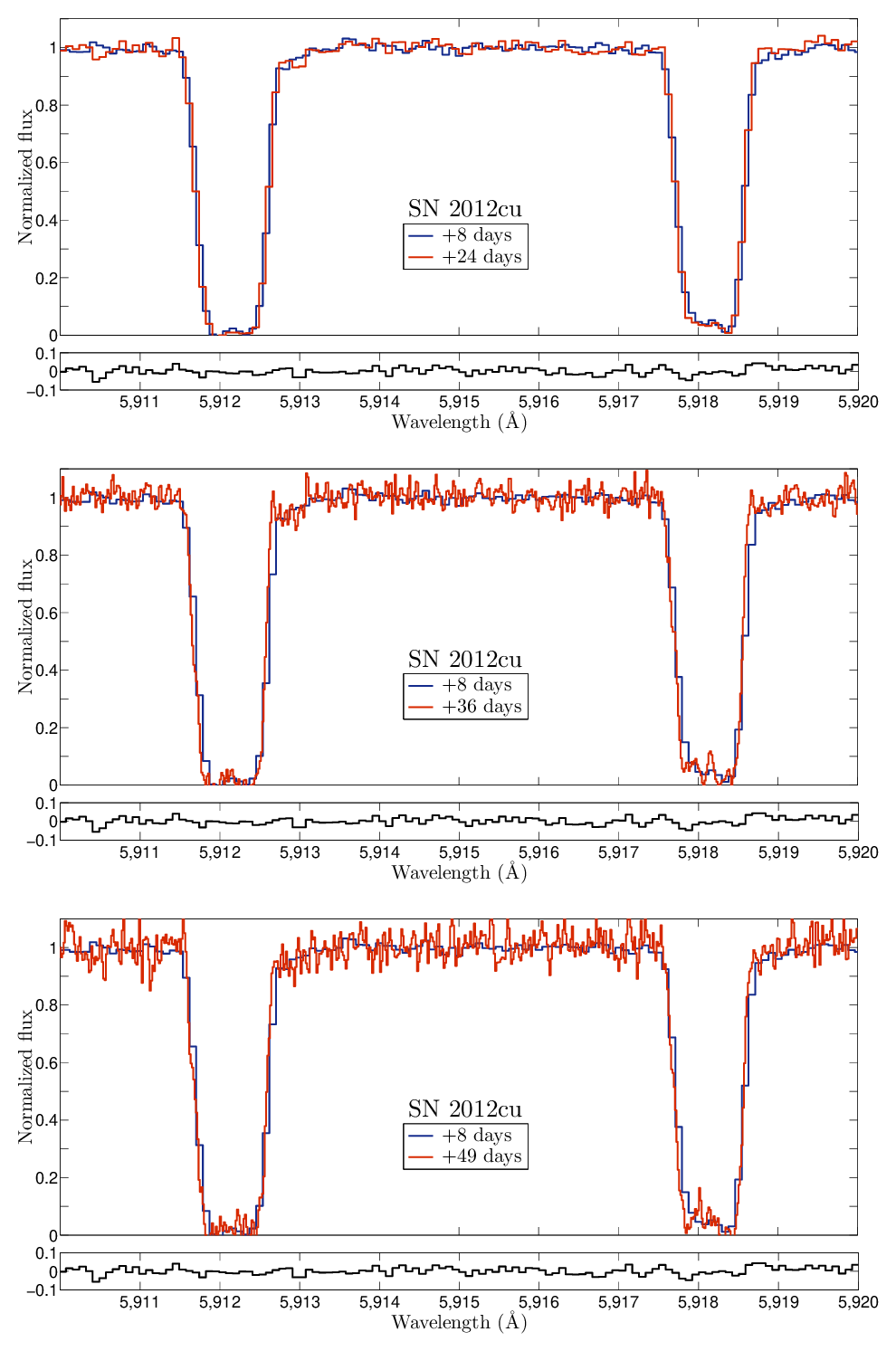}
  \caption{The multi-epoch spectra of SN~2012cu.}
\label{fig:12cu}
\end{figure}

\begin{figure}
  \includegraphics[width=1.0\columnwidth]{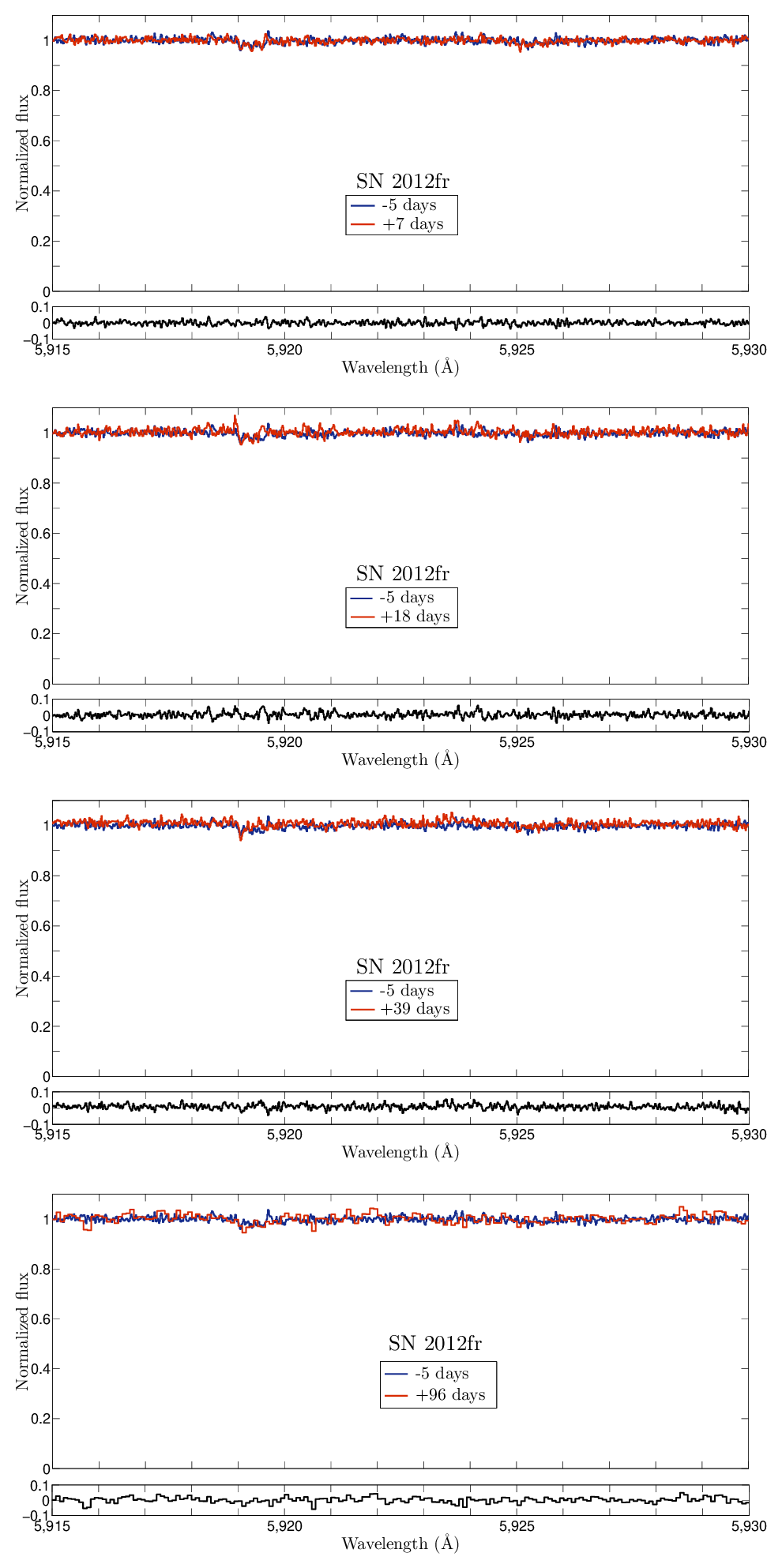}
  \caption{The multi-epoch spectra of SN~2012fr. SN or galaxy ($z=0.005457$) Na~I~D features are expected to appear around 5922\AA\, \& 5928\AA (D$_2$ and D$_1$, respectively). The very weak features around 5919\AA\ are slightly under-subtracted telluric features. No host or SN features are observed.}
\label{fig:12fr}
\end{figure}

\begin{figure}
  \includegraphics[width=1.0\columnwidth]{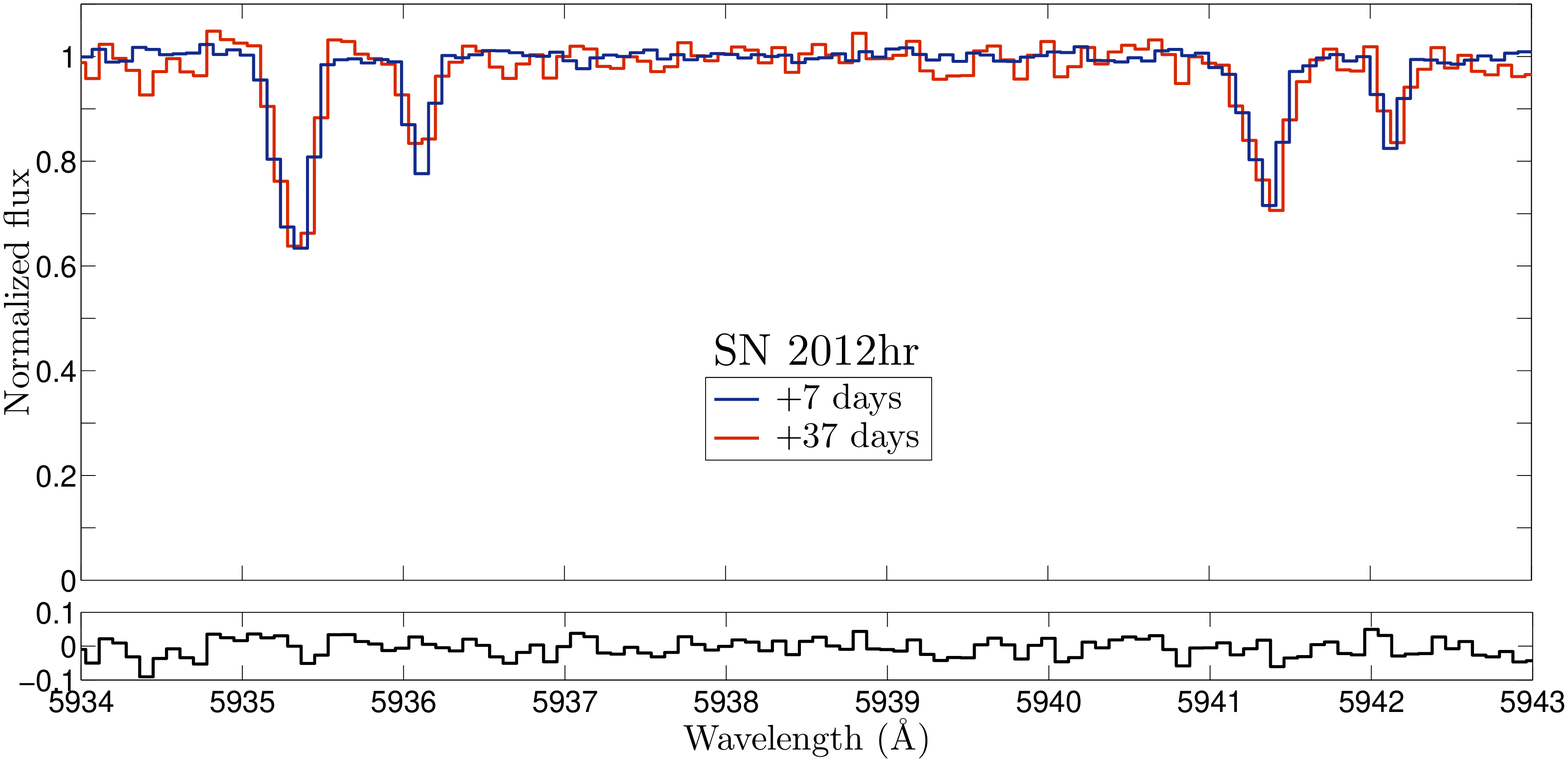}
  \caption{The dual-epoch spectra of SN~2012hr. The observed variability in the D$_2$ red-most feature ($\sim5936.05$\AA) is both not observed in the corresponding D$_1$ line and within the calculated errors.}
\label{fig:12hr}
\end{figure}

\begin{figure}
  \includegraphics[width=1.0\columnwidth]{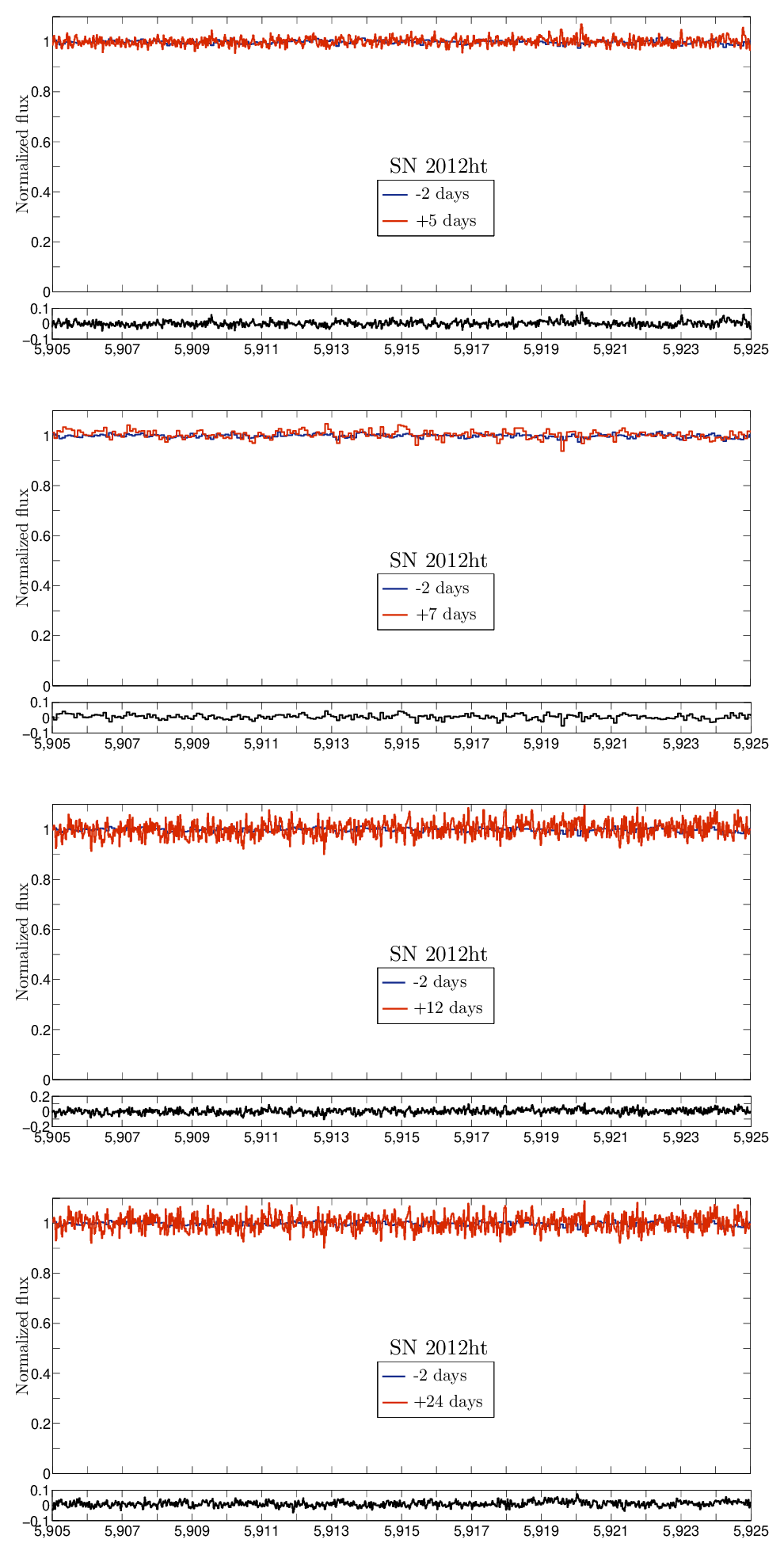}
  \caption{The multi-epoch spectra of SN~2012ht. SN or host galaxy ($z=0.003559$) Na~I~D features are expected to appear around 5911\AA\, \& 5917\AA. No features are observed.}
\label{fig:12ht}
\end{figure}

\begin{figure}
  \includegraphics[width=1.0\columnwidth]{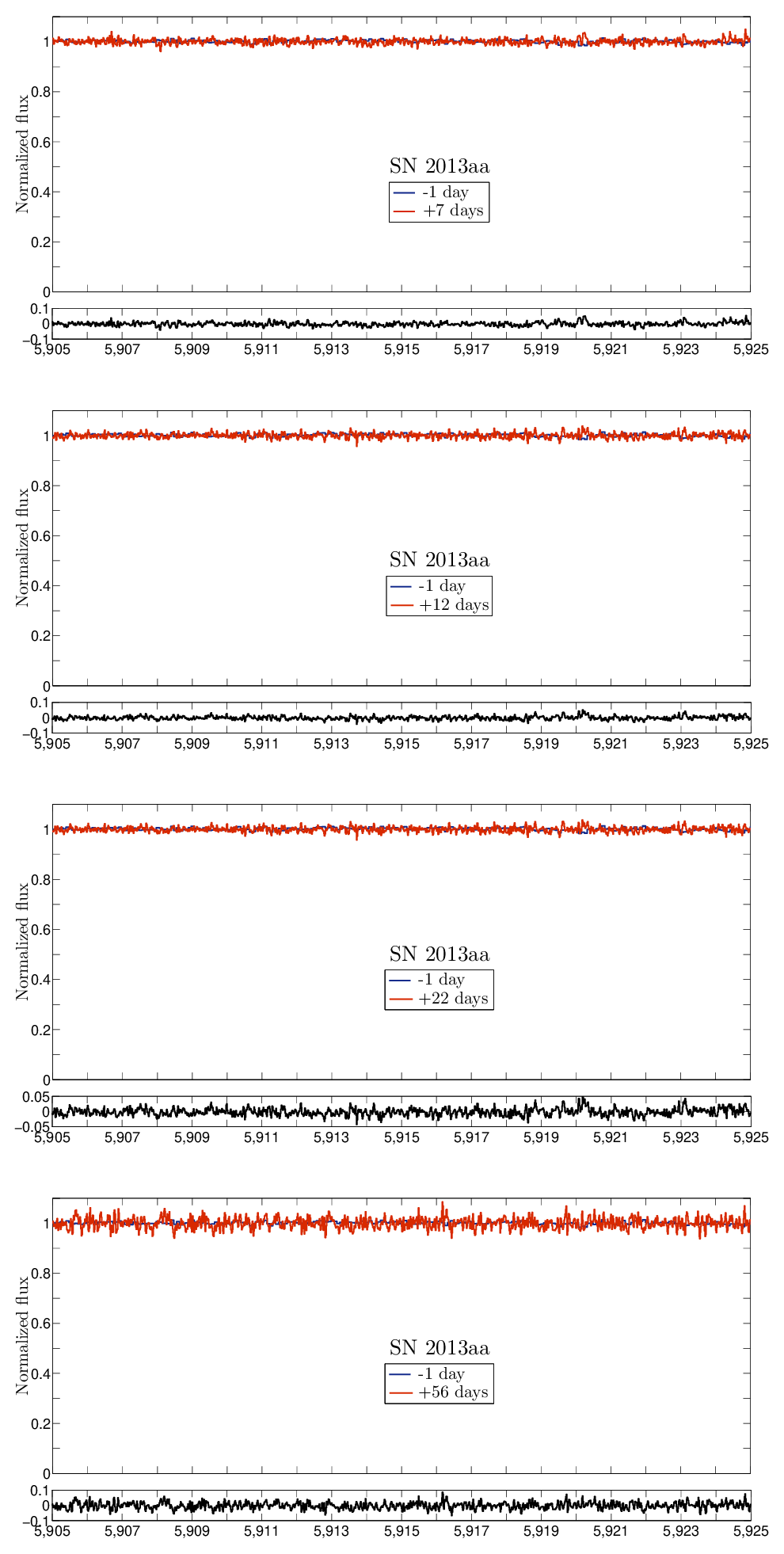}
  \caption{The multi-epoch spectra of SN~2013aa. SN or host galaxy ($z=0.003999$) Na~I~D features are expected to appear around 5913.5\AA\, \& 5919.5\AA. No features are observed.}
\label{fig:13aa}
\end{figure}

\begin{figure}
  \includegraphics[width=1.0\columnwidth]{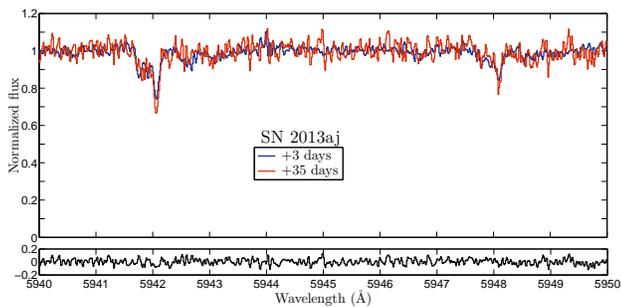}
  \caption{The dual-epoch spectra of SN~2013aj.}
\label{fig:13aj}
\end{figure}

\begin{figure*}
  \includegraphics[width=1.0\textwidth]{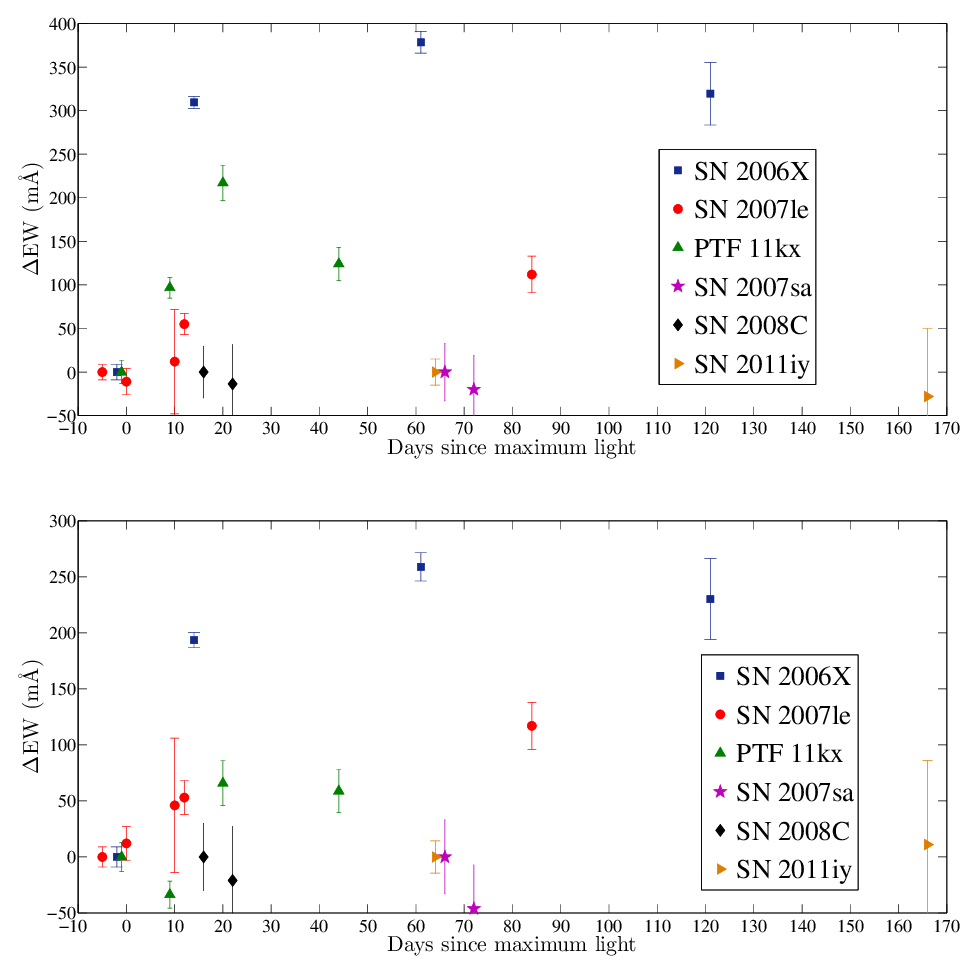}
  \caption{Changes in the EW of previously published events SN~2006X, SN~2007le, and PTF~11kx as a function of days since maximum light. The upper panel presents the change in the D$_2$ line and the lower in the D$_1$ lines. The error bars represent $3\sigma$ errors. the data points for SN~2007sa, SN~2008C, and SN~2011iy are plotted to show that their observations were either performed on epochs too closely spaced apart or that their first epochs were obtained too late.}
\label{fig:EW_old}
\end{figure*}

Using the IRAF RVCORRECT and DOPCOR routines we applied the appropriate heliocentric correction to all the spectra. Using the IRAF SPLOT routine we normalized the spectra, measured the RMS and the signal to noise ratio (S/N) on the continuum in the vicinity of the D$_2$ and D$_1$ lines. We removed telluric features using a synthetic telluric spectrum produced by the Line By Line Radiative Transfer Model \citep{Clough_et_al2005JQSRT91_233C} based on the HITRAN database \citep{Rothman_et_al2009JQSRT110_533R}. For more details see \citet[][their Appendix A]{Patat_et_al2013A&A549_62}. We measured the equivalent width (EW) of the absorption features using the RMS as the {\it sigma0} parameter in the SPLOT error estimation. EW errors ignore continuum placement. Values of EW, error, S/N, and the D$_1$ to D$_2$ ratio for each epoch are given in Table \ref{tab:EW_new}. Comparisons between the first epoch and every subsequent epoch obtained for each SN are presented in Figs. \ref{fig:07on}-\ref{fig:13aj}. First epoch spectra are in blue and subsequent epochs in red. The black line in the narrow panel of each figure is the difference spectrum, i.e., the subtraction of the earlier epoch spectrum from the later one. For comparison we present the EW, error and S/N measurements for SN~2006X, SN~2007le and PTF~11kx in Table \ref{tab:EW_old}. We performed the measurements for SN~2006X and PTF~11kx and used the measurements previously published for SN~2007le \citep{Simon_et_al2009ApJ702_1157}. Fig \ref{fig:EW_old} shows a plot of the EW change as a function of days since maximum light for these three cases, SN~2007on, SN~2007C, and SN~2011iy. \par

\begin{table*}
  \centering
  \caption{Measurements for the observed events.}
  \begin{tabular}{ c c c c c c c c c c }
    & \multirow{2}{*}{Phase\footnotemark[1]}&\multirow{2}{*}{Instrument}&\multicolumn{2}{c}{EW (m\AA)}&\multirow{2}{*}{D$_1$/D$_2$\footnotemark[2]} & \multirow{2}{*}{S/N} \\
    & & & D$_2$ & D$_1$ & & \\
    \hline
    \hline
    \multirow{2}{*}{SN~2007on}& +4 & MIKE & 4$\pm$6 & 0$\pm$5 & N/A & 116 \\
                         & +61 & HIRES & 0$\pm$8 & 0$\pm$6 & N/A & 48 \\
    \hline
    \multirow{2}{*}{SN~2007sa}& +66 & HIRES & 558$\pm$11 & 475$\pm$11 & 0.85 & 21 \\
                         & +72 & HIRES & 538$\pm$13 & 429$\pm$13 & 0.8 & 18 \\
    \hline
    \multirow{5}{*}{SN~2007sr}& +6 & MIKE & 152$\pm$2 & 146$\pm$2 & 0.96 & 109 \\
                         & +7 & MIKE & 152$\pm$2 & 137$\pm$2 & 0.9 & 112 \\
                         & +8 & MIKE & 152$\pm$2 & 138$\pm$2 & 0.91 & 117 \\
                         & +10 & EAE & 152$\pm$4 & 120$\pm$4 & 0.79 & 51 \\
                         & +30 & HIRES & 153$\pm$2 & 126$\pm$2 & 0.82 & 104 \\
    \hline
    \multirow{2}{*}{SN~2008C}& +16 & HIRES &  637$\pm$10 & 444$\pm$10 & 0.7 & 34 \\
                         & +22 & HIRES &  624$\pm$15 & 423$\pm$16 & 0.68 & 22 \\
    \hline
    \multirow{3}{*}{SN~2008dt}& +6 & HIRES &  409$\pm$33 & 287$\pm$33 & 0.7 & 10 \\
                         & +12 & HIRES &  490$\pm$32 & 325$\pm$32 & 0.66 & 11 \\
                         & +18 & HIRES &  430$\pm$25 & 354$\pm$25 & 0.82 & 13 \\
    \hline
    \multirow{2}{*}{SN~2009ds}& +7 & MIKE & 676$\pm$5 & 497$\pm$5 & 0.74 & 74 \\
                         & +24 & MIKE & 704$\pm$6 & 510$\pm$6 & 0.72 & 68 \\
    \hline
    \multirow{3}{*}{SN~2010A}& -8 & MIKE & 453$\pm$3 & 308$\pm$4 & 0.68 & 90 \\
                         & -7 & MIKE & 456$\pm$3 & 284$\pm$4 & 0.62 & 88 \\
                         & +14 & MIKE & 448$\pm$7 & 301$\pm$9 & 0.67 & 37 \\
    \hline
    \multirow{2}{*}{SN~2011iy}& +64 & MIKE & 1314$\pm$5 & 1066$\pm$5 & 0.81 & 69 \\
                              & +166 & MIKE & 1286$\pm$26 & 1077$\pm$25 & 0.84 & 13 \\
    \hline
    \multirow{4}{*}{SN~2012cu}& +8 & MIKE & 928$\pm$6 & 857$\pm$6 & 0.92 & 79 \\
                             & +24 & MIKE & 936$\pm$7 & 851$\pm$7 & 0.91 & 61 \\
                             & +36 & UVES & 918$\pm$6 & 838$\pm$6 & 0.91 & 33 \\
                             & +49 & UVES & 919$\pm$8 & 851$\pm$8 & 0.93 & 24 \\
    \hline
    \multirow{5}{*}{SN~2012fr}& -5 & UVES & 0$\pm$2 & 0$\pm$2 & N/A & 111 \\
                             & +7 & UVES & 0$\pm$2 & 0$\pm$2 & N/A & 119 \\
                             & +18 & UVES & 0$\pm$4 & 3$\pm$4 & N/A & 83 \\
                             & +39 & UVES & 9$\pm$3 & 28$\pm$3 & N/A & 102 \\
                             & +96 & MIKE & 10$\pm$11 & 18$\pm$11 & N/A & 52 \\
    \hline
    \multirow{2}{*}{SN~2012hr}& +7 & MIKE & 133$\pm$3 & 102$\pm$4 & 0.77 & 123 \\
                             & +37 & MIKE & 122$\pm$8 & 120$\pm$9 & 0.98 & 45 \\
    \hline
    \multirow{5}{*}{SN~2012ht}& -2 & MIKE & 0$\pm$4 & 1$\pm$4 & N/A & 148 \\
                             & +5 & UVES & 3$\pm$4 & 5$\pm$4 & N/A & 67 \\
                             & +7 & MIKE & 0$\pm$9 & 0$\pm$9 & N/A & 66 \\
                             & +12 & UVES & 4$\pm$7 & 5$\pm$7 & N/A & 33 \\
                             & +24 & UVES & 1$\pm$3 & 0$\pm$3 & N/A & 70 \\
    \hline
    \multirow{5}{*}{SN~2013aa}& -1 & MIKE & 0$\pm$3 & 0$\pm$3 & N/A & 182 \\
                             & +7 & UVES & 2$\pm$3 & 0$\pm$3 & N/A & 96 \\
                             & +12 & UVES & 1$\pm$2 & 0$\pm$2 & N/A & 101 \\
                             & +22 & UVES & 2$\pm$3 & 1$\pm$3 & N/A & 69 \\
                             & +56 & UVES & 4$\pm$5 & 2$\pm$5 & N/A & 46 \\
    \hline
    \multirow{2}{*}{SN~2013aj}& +3 & UVES & 89$\pm$2 & 57$\pm$2 & 0.64 & 73 \\
                             & +35 & UVES & 83$\pm$6 & 51$\pm$6 & 0.62 & 28 \\
    \hline
\multicolumn{7}{l}{$^1$ In days since max.} \\
\multicolumn{7}{l}{$^2$ For events that do not exhibit sodium lines the ratio is marked as N/A.}
  \end{tabular}
\label{tab:EW_new}
\end{table*}

\begin{table}
  \centering
  \caption{Measurements for the previously published events.}
  \begin{tabular}{ c c c c c c c }
    & Phase (days &\multicolumn{2}{c}{EW (m\AA)}&\multirow{2}{*}{S/N}\\
    & since max) & D$_2$ & D$_1$ & \\
    \hline
    \hline
    \multirow{4}{*}{SN~2006X\footnotemark[1]}& -2 & 828$\pm$3 & 688$\pm$3 & 78 \\
                         & 14 & 1137$\pm$2 & 882$\pm$2 & 101 \\
                         & 61 & 1206$\pm$4 & 947$\pm$4 & 56 \\
                         & 121 & 1147$\pm$12 & 918$\pm$12 & 20 \\
    \hline
    \multirow{5}{*}{SN~2007le\footnotemark[2]} & -5 & 894$\pm$3 & 649$\pm$3 & 113 \\
                         & 0 & 883$\pm$4 & 661$\pm$5 & 71 \\
                         & 10 & 906$\pm$20 & 695$\pm$20 & 16 \\
                         & 12 & 949$\pm$4 & 702$\pm$5 & 76 \\
                         & 84  & 1006$\pm$7 & 766$\pm$7 & 46 \\
    \hline
    \multirow{4}{*}{PTF~11kx\footnotemark[1]$^,$\footnotemark[3]}& -1 & 75$\pm$13 & 124$\pm$13 & 21 \\
                         & 9 & 172$\pm$12 & 91$\pm$12 & 24 \\
                         & 20 & 292$\pm$20 & 190$\pm$20 & 14 \\
                         & 44 & 199$\pm$19 & 183$\pm$19 & 15 \\
    \hline
\multicolumn{5}{l}{$^1$ New measurements.} \\
\multicolumn{5}{l}{$^2$ Taken from \citet{Simon_et_al2009ApJ702_1157}.}\\
\multicolumn{5}{l}{$^3$ Measured only on the variable features.}
  \end{tabular}
\label{tab:EW_old}
\end{table}

In Fig. \ref{fig:07on} we present a comparison between the two spectra of SN~2007on. Based on the host galaxy redshift, $z=0.006494$ \citep{Graham_et_al1998A&AS133_325G}, the SN or host sodium features should have been visible, if present, within the given range. Both epochs seem to be featureless and are  to show that there is no significant difference between them. The difference spectrum does not reveal any significant change between the two epochs. The EW values of the D$_2$ and D$_1$ lines are consistent with zero.  \par

In Fig. \ref{fig:07sa} we present the region of the Na~I~D features observed in SN~2007sa spectra that are consistent with the SN or host galaxy redshift. Both epochs are overplotted to show that there is no significant difference between them. Moreover, the difference spectrum does not reveal any significant change that is not within the noise level. Though the EW values of the D$_2$ and D$_1$ lines show a slight decreasing trend between the first and the second epoch they are still consistent within the calculated errors, $1.2\sigma$ and $2.7\sigma$ for the D$_2$ and D$_1$ lines respectively. Given the small time separation  between the two epochs and the lateness of the first spectrum, we consider it unlikely that these changes are real. \par

In Fig. \ref{fig:07sr} we present the sodium region of the multi-epoch spectra of SN~2007sr. These five epochs were obtained using three different spectrographs - MIKE, EAE, and HIRES. Though some apparent variability can be seen the D$_2$ EW values measured for all epochs are all consistent within the calculated errors. The changes in the D$_1$ measurments in the forth and fifth epochs are close to the $4\sigma$ level but are most likely not real, rather due to normalization of the spectrum near the edge of the D$_1$ line, and they do not appear in the stronger D$_2$ lines. The observed variability can be attributed to the difference between the spectral resolution of the three instruments. \par

An examination of the SN~2008C spectra in Fig. \ref{fig:08C} does not reveal lines that are significantly variable in both lines of the sodium doublet. Though the EW values exhibit a slight decreasing trend they are consistent with one another as they are in agreement within $\sim1\sigma$. In addition, the decrease in EW is counter to the expected behavior from ionization-recombination. Rather, this decrease, if it were significant, would be consistent with ongoing ionization at epochs that are a 16 and 22 days after maximum light. \par

Each panel of Fig. \ref{fig:08dt} shows comparisons between second (upper) or third (lower) epoch and the first epoch of SN~2008dt. Though some lines exhibit a visible change, these changes are all within the variations expected due to noise. This can be seen quite clearly in the difference spectra. Moreover, when we examine the EW values we see that the D$_2$ and the D$_1$ line seem to show different trends. As the sodium D lines are a doublet, a real trend should manifest itself in the same manner in both lines. Moreover, given the error estimates the EW values of the three epochs are consistent with one another well within the $2\sigma$ level. \par

The dual-epochs of SN~2009ds are presented in Fig. \ref{fig:09ds}. The difference spectra given in the lower panel demonstrate that the slight changes between the two epochs are well within the noise level. EW values of the D$_1$ lines are consistent with one another. The EW values of the D$_2$ show a larger difference and an agreement of only $3.6\sigma$. If we look at the EW of the D$_1$ and D$_2$ lines combined we see that the two epoch show a difference of $3.7\sigma$. This is a small but significant detection of time variability. Whether or not this variability can be associated with CSM will be discussed in \S\ref{sec:discussion}. The two epochs are spread over a period that spans the previously detected time-variability phase. \par

Examination of Fig. \ref{fig:10A} does not reveal significant variations in the absorption line profiles observed in the spectra of SN~2010A. All variations are in the D$_2$ lines are in agreement with the S/N and the calculated errors, at the $1\sigma$ level or bellow. However, the EW of the D$_1$ lines of the second epoch does not agree with the first epoch at a $4\sigma$ level. Nevertheless, the EW of the D$_1$ lines in the second and third epoch agree on a $1.7\sigma$ level and between the first and third epoch on a $0.7\sigma$ level. Moreover, if this variability were true, the stronger D$_2$ lines should have shown a grater disagreement, yet they do not. \par

The second epoch of SN~2011iy was obtained with very poor S/N. Even though some variability is seen in Fig. \ref{fig:11iy} the difference spectrum shows that the difference is well within the noise level. The variations in the absorption lines are of the same order as the variations seen in the continuum between the doublet. Moreover, the values of the EW are in agreement within 1.1$\sigma$.

All four epochs of SN~2012cu seem to be well in agreement with one another as seen in Fig. \ref{fig:12cu}. The EW values also agree with each other quite well. The largest disagreement is a $2.3\sigma$ difference between the EW of the D$_1$ features measured for the first and third epochs. The lines are highly saturated. \par

There are some very weak features around 5919\AA\ in the spectra of SN~2012fr (see Fig. \ref{fig:12fr}). These are slightly under-subtracted telluric features. Otherwise the region seems to be featureless on all epochs. SN~2012ht and SN~2013aa also exhibit featureless spectra, see Fig. \ref{fig:12ht} and \ref{fig:13aa} respectively. \par

The spectra of SN~2012hr are presented in Fig. \ref{fig:12hr}. We can see that the comparison between the two epochs does not reveal significant variability. Though the red-most D$_2$ feature shows a slight decrease in intensity this is not observed in the D$_1$ feature. In addition, the EW values of the D$_2$ features are consistent within a $1.3\sigma$ level. \par

The two epochs of SN~2013aj are presented in Fig. \ref{fig:13aj}. While the blue-most feature of the sodium doublet appears to experience some strengthening, the EWs of both epochs agree with each other within the $1\sigma$ level. Moreover, the difference spectrum shows that the differences are within the noise level. \par

To summarize, 13 of the SNe~Ia presented in this study do not exhibit significant time-variable absorption features that could be indicative of the presence of CSM along the line-of-sight to these events. The difference spectra of these events are consistent with zero within the S/N. In contrast, an examination of SN~2006X, SN~2007le, and PTF~11kx, the three previously published events for which time-variability was detected in high-resolution-spectra, clearly reveals the time-variability in the Na~I~D absorption features. One SN, 2009ds, does indeed exhibit variability. This variability is not as profound as the variability exhibited in the previously published detections, and in the next section we will discuss whether this variability is indicative of CSM or whether it is consistent with other explanations.   \par

% ========================================
\section{DISCUSSION} 
\label{sec:discussion}
% ========================================

Comparing the shape and EW of the sodium D line features exhibited in the SN~Ia events presented in this study reveals no significant time-variability that may be associated with CSM in all but one case, SN~2009ds. A question immediately arises what is the cause of this variability. The measured change in EW is $\Delta\rm{EW}_{D_2}=28\pm8$m\AA\, and $\Delta\rm{EW}_{D_1}=13\pm8$m\AA. This change is small compared to the variability observed in SN~2006X, SN~2007le, and PTF~11kx. It is comperable to the observed variability in SN~2011fe, $\Delta\rm{EW}_{D_2}=15.6\pm6.5$m\AA\, \citep{Patat_et_al2013A&A549_62}, which was shown to be consistant with the expected short time-scale variations in interstellar absorptions produced by an expending SN photosphere combined with a patchy ISM \citep{Patat_et_al2010A&A514A_78P}. Therefore, even if the observed variability is real we conclude that it is most likely associated to the ISM and not to the CSM environment of SN~2009ds.  \par

As to whether the reported non-detections are robust. A robust non-detection should have an early first epoch that can serve as a good zero point with which to compare later epochs. As the CSM, if present, begins to recombine after maximum light, the initial epoch should be obtained around maximum light. An initial spectrum obtained later, might have been obtained after some, or all, of the CSM has recombined, and will not serve as a good zero point for comparison.  \par

The first spectra of both SN~2007sa and SN~2011iy were obtained more then 60 days after maximum light. Based on previous CSM detections we expect that by this phase any CSM that was ionized by the UV flash will have already re-combined (see Fig. \ref{fig:EW_old}). Therefore, such late spectra cannot rule out variability at earlier stages and therefore should not be used as a non-detection and should not be included in statistical analysis of multi-epoch high-spectral-resolution samples of SNe~Ia. \par

Based on previous cases (SN~2006X, SN~2007le, and PTF~11kx) we cannot determine that any variability would have been seen between the two closely-spaced epochs we have for SN~2008C (days +18 and +22; see Fig. \ref{fig:EW_old}). So, we argue that this non-detection should not be considered robust. This case emphasizes the need for an early-time first epoch and at least on more well spaced epoch. \par

The first epochs of the remaining 11 events were obtained no later then 8 days after maximum light. If we disregard epochs of SN~2006X, SN~2007le, and PTF~11kx that were obtained earlier then 8 days we can still observe significant time-variability in all three events. Therefore, we argue that the non-detection reported for these events are robust. This criterion will have to be revised if future observations will show events for which all the time-variability occurs earlier then 8 days after maximum light. \par

The D$_1$/D$_2$ ratio for many of our events is considerably larger then 0.5, the ratio for optically thin gas. In the optically thick regime the line profiles are less sensitive to changes in the column density, making them harder to identify, especially in low S/N spectra. Nevertheless, the obvious fact that we do not observe time variability in the spectra of these events at these epochs is an indication that the environment along the line-of-sight to these events is different with respect to that along the line-of-sight to SN~2006X, SN~2007le, and PTF~11kx. \par

In addition, we cannot rule out the presence of weak absorption lines that will be undetectable due to the noise in the spectra. Following \citet{Leonard&Filippenko2001PASP113_920L}, one can quantify the upper detection limit when no features are apparent in the spectrum as, 
\begin{equation}
W_\lambda(3\sigma)=3\Delta\lambda\Delta I\sqrt{\frac{W_{line}}{\Delta\lambda}}\sqrt{\frac{1}{B}},
\end{equation}
where $\Delta\lambda$ is the spectral resolution, $\Delta I$ the root-mean-square fluctuations of the normalized flux, W$_{line}$ is the width of the feature, and B is the number of bins per resolution element. Assuming that these lines are optically thin we can follow \citep[][\S 3.4.c]{Spitzer1978ppim_book} and use the curve-of-growth to convert equivalent width into column densities using,
\begin{equation}
N=\frac{m_e c^2}{\pi e^2}\frac{W_\lambda}{f\lambda^2}=1.13\times10^{20}\frac{W_\lambda}{f\lambda^2},
\end{equation}
where $f$ is the oscillator strength. Given the different spectral resolution of the different instruments and the S/N of the different observations, and assuming $W_\lambda=0.2$\AA, we can conclude that any feature arising from a Na~I column density of more then a few times $10^{10}$ cm$^{-2}$ should have been detectable in the spectra of the majority of these events. The extreme cases are SN~2012fr , for which the detectability limit is $\sim7.8\times10^{9}$ cm$^{-2}$ and SN~2007on and SN~2008dt, for which the limits are $6\times10^{10}$ cm$^{-2}$ and $8\times10^{10}$ cm$^{-2}$, respectively. Therefore, a line-of-sight environment similar to those observed for SN~2006X, SN2007le, and PTF~11kx should have produced detectable lines. Assuming all the sodium is neutral and a solar abundance ratio, $\log Na/H=-6.3$, we get a hydrogen column density limit of $\sim10^{17}$ cm$^{-2}$. Assuming this material is distributed in a thin spherical shell at radius $R$ the mass of the shell will be,
\begin{equation}
M_{shell}\simeq\left(\frac{R}{10^{17}\rm{cm}}\right)^2\times10^{-5}\,\rm{M}_\odot,
\end{equation}
though this estimate is only a rough estimate and most likely an over estimate. For SN~2012fr the limit is approximately one order of magnitude lower. \par

Our spectroscopic observations only probe the CSM along the line-of-sight, therefore, we cannot rule out presence of material off the line-of-sight. Nevertheless, simulations of proposed progenitor systems \citep[e.g.,][]{MohamedBoothPodsiadlowski2013ASPC469_323, ShenGuillochon&Foley2013ApJ770L_35, Raskin&Kasen2013ApJ772_1} may offer an insight to whether they can form CSM in a way that allows for line-of-sight environments that are consistent with our observations, especially the relatively clean featureless line-of-sights observed toward SN~2007on, SN~2012fr, SN~2012ht, and SN~2013aa. Such a comparison between theoretical results and the observed multi-epoch high-spectral-resolution sample can be highly insightful, but remains beyond the scope of this paper.

% ========================================
\section{CONCLUSIONS} 
\label{sec:conclusions}
% ========================================
Our data extends the published multi-epoch high-spectral-resolution sample size to 17 events for which robust detection or non-detection can be claimed. Three events exhibit time-variable Na~I~D features attributed to CSM. Assuming Poisson statistics we find that $18\%\pm11\%$ ($19\%\pm12\%$ if we exclude SN~2007on which occurred in an elliptical host) of the events in the enlarged sample exhibit time-variable features that can be associated with CSM. This result is in agreement with the results of S11 and \citet{Maguire_et_al2013accepted} who estimated the fraction to be $20-25\%$ of nearby SNe~Ia in late-type galaxy hosts. This estimate is a lower limit as CSM may lay off the line-of-sight or on the line-of-sight but further away from the progenitor system. In the first case this material will not be visible in the spectrum. In the later, the CSM lines will not vary with time and be regarded as a non-detection. Moreover, though this is the largest multi-epoch high-spectral-resolution SN~Ia sample to date it is still relatively small. A larger sample will be useful to shed light on the ratio between SNe~Ia with CSM and those without. This ratio and the study of the properties of detected CSM can help us to disentangle the different proposed channels leading to SNe~Ia explosions. In order for the published multi-epoch high-spectral-resolution sample to be more complete and not biased toward events with detected time-variability it is critical to publish all the observed events. With the publication of this data set the sample of SNe~Ia observed by the Keck-Magellan effort till 2009 is complete. The VLT 2008--2009 sample is still being worked on and will be soon published (Cox et al., in preparation). \par

%================================================================

\section*{Acknowledgments}

A.S. is supported by a Minerva Fellowship. The research of A.G. is supported by the EU/FP7 via an ERC grant no. 307260, the Minerva ARCHES prize and the Kimmel award. \\
The authors would like to acknowledge the generosity of the late Wallace L.~W.~Sargent in providing data. The authors would also like to acknowledge the help of D.~J.~Osip and J.~F.~Steiner in obtaining the data. \\
Partially based on observations made with ESO Telescopes at the La Silla Paranal
Observatory, Chile, under programme ID 289.D-5023, 290.D-5010, 290.D-5023, 091.D-0780. \\
This paper includes data gathered with the 6.5 meter Magellan Telescopes located at Las Campanas Observatory, Chile. \\
Some of the data presented herein were obtained at the W.M. Keck Observatory, which is operated as a scientific partnership among the California Institute of Technology, the University of California and the National Aeronautics and Space Administration. The Observatory was made possible by the generous financial support of the W.M. Keck Foundation. The authors wish to recognize and acknowledge the very significant cultural role and reverence that the summit of Mauna Kea has always had within the indigenous Hawaiian community.  We are most fortunate to have the opportunity to conduct observations from this mountain.

\footnotesize{
  \bibliographystyle{mn2e}
  \bibliography{mybib}
}

\appendix
\section{Description of the observations}\label{app:obs_desc}

In this appendix we provide information regarding the discovery and spectroscopic observations of each SN in the presented sample. \\

%\subsection{SN~2007on}
%\label{data:07on}
\noindent{\bf SN~2007on} was discovered on 2007-11-05.25 UT (all dates in this paper are given in UT) 12" west and 68" north of the center of the elliptical galaxy NGC 1404  \citep{Pollas&Klotz2007CBET1121} and reported to be a young SN~Ia on 2007-11-06 \citep{GalYam_et_al2007ATel1263}. Based on the Carnegie Supernova Project photometric data\footnote{http://csp.obs.carnegiescience.edu/data/lowzSNe/SN2007on/} SN~2007on maximum light in B band occurred around 2007-11-16. HIRES spectra of SN~2007sa were obtained on 2007-11-20.3 and 2008-01-16.2. The first observation was performed 4 days after B band maximum light. \\

%\subsection{SN~2007sa}
%\label{data:07sa}
\noindent{\bf SN~2007sa} was discovered on 2007-11-21.56  1".8 west and 7".7 north of the nucleus of NGC 3499  \citep{MostardiLi2007CBET1161} and reported to be a SN~Ia one month after maximum light on 2007-12-12.14 \citep{Angoletto2007CBET1163}. HIRES observations of SN~2007sa were performed on 2008-01-17.4 and 2008-01-23.59. Both spectroscopic epochs were obtained quite late (the first one 66 days after maximum light) and fairly close together, reducing the sensitivity to variable absorption. \\

%\subsection{SN~2007sr}
%\label{data:07sr}
\noindent{\bf SN~2007sr} was discovered on 2007-12-18.53 on the southern arm of the antennae galaxies emanating from NGC 4038 \citep{Drake_et_al2007ATel1337}, at 4 days after maximum light \citep{Pojmanski2008CBET1213}. SN~2007sr was classified as a SN~Ia on 2007-12-19 \citep{Naito_et_al2007CBET1173}. SN~2007sr was observed using MIKE on 2007-12-20, 2007-12-21, and 2007-12-22, using EAE on 2007-12-24, and using HIRES on 2008-01-17. The first epoch was obtained around 6 days after maximum light. \\

%\subsection{SN~2008C}
%\label{data:08C}
\noindent{\bf SN~2008C} was discovered on 2008-01-03.27 2".95 west of the center of UGC 3611 \citep{Puckett2008CBET1195} and confirmed as a SN~Ia near maximum light on 2008-01-04.8 \citep{AyaniYamaoka2008CBET1197}. SN~2008C was observed using HIRES on 2008-01-17.64 and 2008-01-23.55. SN~2008C was spectrally typed as a normal Type Ia \citep{Stritzinger_et_al2011AJ142_156}. The first epoch observation was performed 16 days after maximum light \citep{Foley_et_al2012ApJ_752_101}. \\

%\subsection{SN~2008dt}
%\label{data:08dt}
\noindent{\bf SN~2008dt} was discovered on 2008-06-30.33 1".0 east and 5".5 south of the nucleus of NGC 6261 \citep{MadisonLiFilippenko2008CBET1423} and confirmed as a SN~Ia around one week before maximum light on 2008-07-01.21 \citep{BlondinBerlind2008CBET1424}. SN~2008dt was observed using HIRES on 2008-07-06.48, 2008-07-12.48 and 2008-07-18.46. The first epoch observation was performed 6 days after maximum light \citep{Foley_et_al2012ApJ_752_101}. \\

%\subsection{SN~2009ds}
%\label{data:09ds}
\noindent{\bf SN2009ds} was discovered on 2009-04-28.56 12" west and 3" north of the center of NGC 3905 \citep{Nakano_et_al2009CBET1784} and confirmed as a normal SN~Ia around one week before maximum light on 2009-04-29.6 \citep{ChallisCalkins2009CBET1788, Anderson2009CBET1789}. MIKE observations of SN~2009ds were performed on 2009-05-15.96 and 2009-06-01.97. The first epoch observation was performed around 7 days after maximum light. \\

%\subsection{SN~2010A}
%\label{data:10A}
\noindent{\bf SN~2010A} was discovered on 2010-01-04.14 2".4 east and 6".9 north of the center of UGC 2019 \citep{Cox_et_al2010CBET2109}, and classified as a normal Type Ia at 9 days before maximum brightness on 2010-01-07 \citep{Foley&Esquerdo2010CBET2112}. SN~2010A was observed with MIKE on 2010-01-08, 2010-01-09, and 2010-01-30. The first epoch spectrum was obtained about 8 days before maximum light. \\

%\subsection{SN~2011iy}
%\label{data:11iy}
\noindent{\bf SN~2011iy} was discovered on 2011-12-09.86 16".6 east and 6".1 south of the center of NGC 4984 \citep{Itagaki_et_al2011CBET2943}, and classified as a Type Ia at 12 days after maximum light on 2011-12-14.81 \citep{Yamanaka_et_al2011CBET2943}. SN~2011iy was observed with MIKE on 2012-02-04.26 and 2012-06-06.12. The first epoch spectrum was obtained about about 64 days after maximum light. \\

%\subsection{SN~2012cu}
%\label{data:12cu}
\noindent{\bf SN~2012cu} was discovered on 2012-06-11.1 3".1 east and 27".1 south of the nucleus of NGC 4772 \citep{Itagaki_et_al2012CBET3146}, and was classified as a SN~Ia 7 days before maximum light \citep{Marion_et_al2012CBET3146, Zhang_et_al2012CBET3146}. MIKE observations were performed on 2012-06-30.01 and 2012-07-16.03. UVES observations were performed on 2012-07-28.99 and 2012-08-10.98. The first epoch observation was performed around 8 days after maximum light. \\

%\subsection{SN~2012fr}
%\label{data:12fr}
\noindent{\bf SN~2012fr} was discovered on 2012-10-27.05 3" west and 52" north of the nucleus of NGC 1365 \citep{Koltz_et_al2012CBET3275}, and classified as a Type Ia on 2012-10-28.53 \citep{Childress_et_al2012CBET3275}. Observations with UVES were performed on 2012-11-07.16, 2012-11-19.08, 2012-12-02.13, and 2012-12-21.05, and with MIKE on 2013-02-16.01. The first epoch spectrum was obtained around about 5 days before maximum light. \\

%\subsection{SN~2012hr}
%\label{data:12hr}
\noindent{\bf SN~2012hr} was discovered on 2012-12-16.533 2".3 west and 93".6 north of the center of ESO 121-26 \citep{Drescher_et_al2012CBET3346}. SN~2012hr was classified as a SN~Ia approximately 1 week before maximum light on 2012-12-20.2 \citep{Morrel_et_al2012ATel4663}. MIKE observations were performed on 2013-01-02 and 2013-02-01. The first epoch spectrum was obtained about 7 days after maximum light. \\

%\subsection{SN~2012ht}
%\label{data:12ht}
\noindent{\bf SN~2012ht} was discovered on 2012-12-18.77 19" west and 16" north of the center of NGC 3447 and classified as a SN~Ia 7 days before maximum light on 2012-12-20.4 \citep{Yusa_et_al2012CBET3349}. Based on Swift lightcurves maximum light occurred around 2013-01-04\footnote{http://people.physics.tamu.edu/pbrown/SwiftSN/SN2012ht\_lightcurve.jpg}. Observations with MIKE were performed on 2013-01-02.29 and 2013-01-11.33, and with UVES on 2013-01-09.34, 2013-01-16.34, and 2013-01-28.28. The first epoch observation was performed around 2 days before maximum light. \\

%\subsection{SN~2013aa}
%\label{data:13aa}
\noindent{\bf 2013aa} was discovered 2013-02-13.62 74" west and 180" south of the center of NGC 5643 \citep{Parker_et_al2013CBET3416}. SN~2013aa was classified as a Type Ia a few days before maximum light on 2013-02-15.38 \citep{Parrent_et_al2013ATel4817}. SN2013aa was observed with MIKE on 2013-02-16.38 and with UVES on 2013-02-24.31, 2013-03-01.31, 2013-03-11.39, and 2013-04-14.3. The first epoch spectrum was obtained about 1 days before maximum light. \\

%\subsection{SN~2013aj}
%\label{data:13aj}
\noindent{\bf 2013aj} was discovered on 2013-03-03.14 5".7 east and 8" north of the center of NGC 5339 and was classified on the same night as a SN~Ia 7 days before maximum light \citep{Brimacombe_et_al2013CBET3434}. Observations with UVES were performed on 2013-03-13.29 and 2013-04-14.24. The first epoch observation was performed around 3 days after maximum light.

\label{lastpage}

\end{document}